\pdfoutput=1
\documentclass{webofc}

\usepackage[varg]{txfonts}   

\usepackage{float} 

\usepackage{graphicx}

\usepackage{lineno}

\usepackage{subfigure}
\usepackage{bm}
\usepackage[percent]{overpic}
\usepackage{color}

\usepackage{hyperref}
\usepackage[all]{hypcap}
\hypersetup{
    colorlinks,
    citecolor=blue,
    filecolor=blue,
    linkcolor=black,
    urlcolor=blue
}

\def\pt{p_{\rm T}}

\def 	\bcor{b_{\rm corr}}
\def 	\C2{C_{\rm 2}}

\def\av#1{\langle #1 \rangle}

\def\sNN{\mbox{$\sqrt{s_{_{\rm NN}}}$}}

\def\pf{\overline{p}^{F}}

\def\vf{\varphi}

\def\rF{r^F}
\def\rB{r^B}

\def\nK{n_{\rm K}}
\def\npi{n_{\rm \pi}}

\def\nKF{n_{\rm K}^F}
\def\npiF{n_{\rm \pi}^F}
\def\nKB{n_{\rm K}^B}
\def\npiB{n_{\rm \pi}^B}

\def\naF{n_{  a}^F}
\def\nbF{n_{  b}^F}
\def\naB{n_{  a}^B}
\def\nbB{n_{  b}^B}

\def\nuFB{\nu_{\rm FB}}

\begin{document}
\title{
Recent developments in particle yield fluctuation \\
measurements
}
%
%


\author{\firstname{Igor} \lastname{Altsybeev}\inst{1}\fnsep\thanks{
\email{i.altsybeev@spbu.ru}}
}

\institute{Saint-Petersburg State University, Saint Petersburg, Russia      }

\abstract{In relativistic heavy-ion collisions, 
properties of the initial state
and effects arising during  
evolution of the medium,
such as a transition between the
hadronic and partonic phases, 
should  reflect themselves in event-by-event fluctuations
of the number of produced particles.
In this paper, 
recent measurements of
several event-by-event observables, 
namely,
dynamical fluctuations of relative particle yields 
and  forward-backward correlations of different types, 
are discussed.
Also, new observables for forward-backward correlation studies are proposed:
correlations between 
ratios of identified particle yields
 in two separated acceptance intervals 
and
the correlation between the  ratio in one interval and 
average transverse momentum in another.
}

\maketitle

\section{Introduction}
\label{intro}

Measurement of the properties of dense nuclear matter and search for
 signs of its transition to the state of deconfinement
are performed by studying collisions between heavy ions. 
One way to probe the 
phase structure of strongly interacting matter 
is the analysis of particle production in the framework of
the statistical hadronization model, which accounts for the thermal distribution of particle species.
At the LHC energies, where 
baryo-chemical potential is close to zero,
results of this model can be directly compared with the lattice QCD
calculations.
The value of the pseudo-critical temperature 
extracted from data 
is $T_c=156.2 \pm 1.5$ MeV \cite{PBM_et_al_Nature_StatHadr},
which  corresponds to the lattice results.

Another approach to the search 
 of critical behavior in hadronic
 matter is 
analysis of  fluctuations of various quantities.
Examples of such quantities are 
the conserved charges, in particular, the residual electric charge, 
 baryon number and  strangeness. 
Event-by-event fluctuations help to characterize the properties of the ``bulk'' of the system
and also are closely related to dynamics of the phase transitions.
A non-monotonic behaviour with experimentally varied parameters such as the collision energy, centrality, system size, rapidity is expected \cite{stephanov_et_al_1999}.

In comparison with the ``event-averaged'' observables (like average particle yields, spectra),
observables based on fluctuations are much more sensitive to different biases
which can distort the results:
limited detector acceptance, non-flat  particle registration efficiency
and its dependence on detector occupancy,
contamination by secondary particles, mis-identification of particle species,
conservation laws, 
resonance decays, and, finally,
trivial event-by-event fluctuations of collision geometry  (so-called
``volume fluctuations'').
From this long list of caveats it becomes evident, that for this kind of studies
it is not enough just to define a new observable -- 
it's essential to know how robust this observable is in a real experiment,
and a correction procedure must be provided, if necessary.

In this article,
properties and experimental measurements of 
some fluctuation-based observables are reviewed. 
After that, new fluctuation observables 
based of particle number ratios are introduced.

\section{  Robust observables for fluctuation measurements } 
\label{sec-2}
\vspace*{-0.1cm}

Particle number 
 fluctuations can be quantified by
different measures.
The simplest  one is the variance of multiplicity distribution,
however, 
it depends on the volume 
of the system and its fluctuations
(an {\it extensive}  observable), 
which makes it tricky
 to extract the ``dynamical'' contribution, in which we are really interested in.
Within the grand canonical ensemble,
the volume cancels if the variance is scaled with the mean multiplicity
(an {\it intensive} observable),  
but the dependence 
on volume fluctuations still remains.

Fluctuations can be studied also in terms of variations
in the ratio of 
 particle yields of two distinct species.
Denote number of particles of type $a$ and $b$, 
measured in a given event, as $n_a$ and $n_b$,
then the ratio of the two yields is $r=n_a/n_b$.
The quantity we are interested in is
the variance of this ratio, $\av{\Delta r^2}$.
If one normalizes this quantity to the mean and 
assumes that  deviations of the yields from the means $n_a-\av{n_b}$
and $n_a-\av{n_b}$ are small, 
then the following approximation is valid:
\vspace{-0.12cm}
\begin{equation}
\nu \equiv
\frac{ \av{\Delta r^2}  }{ \av{r}^2  }
\approx \bigg< \bigg(
\frac{ n_a  }{ \av{n_a}  } - \frac{ n_b  }{ \av{n_b}  }
 \bigg)^2
\bigg> \, .
\vspace{-0.12cm}
\end{equation}
In the limit of independent particle production (Poisson statistics),
  this quantity is reduced to just  $\nu_{stat} =1/\av{n_a}+ 1/\av{n_b}$.
In experimental studies,
this trivial contribution is usually  subtracted from the value of $\nu$,
so the observable which is actually measured  is   \cite{nu_dyn_2002}
\vspace{-0.12cm}
\begin{equation}
\label{expr_nudyn}
\nu_{dyn} \equiv \nu - \nu_{stat} =
\frac{\av{n_a(n_a-1)}}{\av{n_a}^2}
+ \frac{\av{n_b(n_b-1)}}{\av{n_b}^2}
- 2\frac{\av{n_a n_b}}{\av{n_a}\av{n_b}} \, .
\vspace{-0.12cm}
\end{equation}
The value of $\nu_{dyn}$ is non-zero when production of particles $A$ and $B$ is correlated (non-Poissonian).
This observable is robust against efficiency losses and volume fluctuations.
It was used to measure
net-charge fluctuations 
in a number of experiments, 
as well as
relative particle yield fluctuations (K/$\pi$, p/$\pi$ and K/p),
for instance, by ALICE \cite{ALICE_rel_part_yield_fluct}. 
It was pointed, that for interpretation of the results, obtained with $\nu_{dyn}$,
the acceptance coverage is crucial, and also that resonance contributions should be better understood before making conclusions about dynamical effects.
It should be stressed that measurements of $\nu_{dyn}$ 
are typically performed within one single pseudorapidity window, 
in most cases this is actually a full $\eta$-acceptance of a given experiment.

A class of observables which do not depend neither on the volume fluctuations
of the system nor on the volume itself (the  so-called {\it strongly intensive} quantities) was introduced in \cite{strongly_int_2011}. 
In fact, one of these observables, denoted as $\Sigma$,
is closely related to $\nu_{dyn}$ -- the difference is 
only in the scaling factor proportional 
to multiplicity densities, which are easy to correct for detector efficiency.
In the model of independent sources, the $\Sigma$ quantity
measures the properties of a single source
\cite{ea_vv_sigma}. 
For Poissonian particle production, $\Sigma=1$.
Recently,
the $\Sigma$ observable 
was used to quantify 
the strength
of the forward-backward multiplicity correlations of charged particles between the two separated
$\eta$-intervals  in Pb--Pb collisions $\sNN=$2.76  TeV \cite{Sigma_ALICE_IS}.
Usage of this observable allows to eliminate 
the big problem of the forward-backward correlation studies -- 
the contribution from the volume fluctuations.
Peculiar change of $\Sigma$ with centrality was obtained,
which is not understood so far.

Another type of forward-backward observable
which is robust against the volume fluctuations
 is a correlation between mean transverse momenta measured event-by-event 
in the two $\eta$-windows.
Non-trivial centrality evolution of the mean-$\pt$ correlation strength   
was recently obtained  in Pb-Pb collisions in ALICE  \cite{FB_mean_pt_ALICE}.

\vspace*{-0.2cm}

\section{ Correlations between ratios of particle yields  }
\vspace*{-0.1cm}


In this section, 
 a new type of observables for correlation studies is introduced, namely, a correlation between 
particle number ratios measured in two separated rapidity intervals.
Note that we can straightforwardly add  splitting of the acceptance intervals 
also in other dimensions (in $\vf$ 
and in $\pt$).
Correlation coefficient between particle number ratios can be written as
\begin{equation}
\label{corr_coeff}
\bcor= \frac{ \av{\rF\cdot \rB}}{  \av{\rF} \av{\rB} } -1 \, ,
\end{equation}
where 
$\rF =    n^F_a /  n^F_b $ and $\rB =    n^B_a /  n^B_b$
are  ratios  of multiplicities for particles of  species $a$ and $b$
measured in the two acceptance windows 
(let us call them ``forward'' and ``backward"), 
and angular brackets denote averaging over events.
For example,
we can consider a ratio of a number of kaons to a number of pions,
$r =    n_K/ n_\pi$, 
in this case the correlator in \eqref{corr_coeff} becomes
$\av{\rF\cdot \rB} =  \bigg<{\nKF}/{\npiF} \cdot  {\nKB}/{\npiB} \bigg>$.

When  multiplicities  of produced particles are large (in semi-central and central 
A-A collisions), 
the event-by-event fluctuations of them 
are expected to be small  relative to the average values.
In this case, it can be shown that the following approximation for the $\bcor$ is valid:
\begin{equation}
\label{beta_approx}
\bcor \approx 
\frac{\av{\naF\naB}}{\av{\naF}\av{\naB}} +\frac{\av{\nbF\nbB}}{\av{\nbF}\av{\nbB}}
-\frac{\av{\naF\nbB}}{\av{\naF}\av{\nbB}} -\frac{\av{\nbF\naB}}{\av{\nbF}\av{\naB}}  
\stackrel{\rm def.}{\equiv} \nuFB^{  a/b} \, .
\end{equation}
Each term in this expression is a normalized correlator between multiplicities in forward and backward intervals.
If we adopt here the notion of the normalized cumulants $R_2^{a b}=\av{n^F_a n^B_b}/\av{n^F_a}\av{n^B_b} - 1$, 
the expression \eqref{beta_approx}  can be rewritten as 
\begin{equation}
\label{beta_approx_using_R2}
\nuFB^{a/b} = R_2^{ aa} + R_2^{bb} - R_2^{ ab} - R_2^{ba} \,.
\end{equation}
In case of independent (Poissonian) particle production 
$\nuFB=0$. 
It is zero also if  only short-range correlations are present in the system, 
and a separation between the two observation windows is large enough to suppress them.
Another important property of $\nuFB$ is its robustness to volume fluctuations 
as well as to  efficiency
of particle detection (if the efficiency is constant within the acceptance windows).
In the model with independent particle-emitting sources, the scaling of this observable 
with the average number of sources $\av{N_{\rm s}}$ 
is $\nuFB=1/\av{N_{\rm s}} \cdot \nuFB^{\rm s}$, where $\nuFB^{\rm s}$ is the value of this quantity 
for a single source.
In fact,  the properties of the  $\nuFB$ observable mentioned above
are 
similar to those of the $\nu_{\rm dyn}$ \eqref{expr_nudyn}
\cite{nu_dyn_2002}.
It is important to note also, that it's possible to correct individual terms of $\nuFB$ (moments of the first and the second order)
for particle mis-identification
by applying the Identity Method \cite{Identity_2011}.

The definition of $\nuFB$  
is also closely connected to observables used for the charge-dependent correlation
studies: 
if  one takes for the analysis
the ratios of multiplicities of positively and negatively charged particles 
(i.e. $\rF =    n^F_+ /  n^F_-$ and $\rB =    n^B_+ /  n^B_-$),
then 
\eqref{beta_approx_using_R2} 
reads as 
$\nuFB^{+/-}= R_2^{++} + R_2^{--} - R_2^{+-} - R_2^{-+}$,
and the relation to the charge-dependent correlation function $R_{\rm CD}$, adopted, for example, in \cite{R2_P2_ALICE}, is
$\nuFB = -4 \cdot R_{\rm CD}$. 
Note that this relation between observables can be treated  in the opposite way:
in case of large enough multiplicities in $F$ and $B$ acceptance windows,
the  $R_{\rm CD}$ observable for charge-dependent correlations is nothing 
else then the approximation for the correlation coefficient $\bcor$ 
 between particle number ratios \eqref{corr_coeff}, multiplied by $-1/4$.
The connection to the balance function observable is also straightforward: 
$b.f. = -  \rho_1^{ch}/4 \cdot \nuFB^{+/-}$, where $b.f.$ is the balance function
between $F$ and $B$ windows and $\rho_1^{ch}$ is a charged particle density.

Consider 
the particular case for the $\bcor$ observable,
when the ratio between number of kaons to number of pions is taken:
$\rF =   \nKF / \npiF$
and
$\rB =  \nKB / \npiB$, then
\begin{equation}
\label{beta_approx_using_R2_K_pi}
   \bcor^{\rm K/\pi} \approx \nuFB^{\rm K/\pi} = R_2^{\rm KK} + R_2^{\pi\pi} - R_2^{\rm K\pi} - R_2^{\rm \pi K} \,.
\end{equation}
With this observable,
a correlation between strangeness production at large $\eta$ gaps can be probed
-- the physics case of interest for thermal models \cite{PBM_et_al_Nature_StatHadr},
models with interactions between color ropes (quark-gluon strings) 
\cite{SFM_strangeness, Bierlich_color_ropes}, etc. 

\begin{figure}[t]
\centering
\begin{overpic}[width=0.48\textwidth]{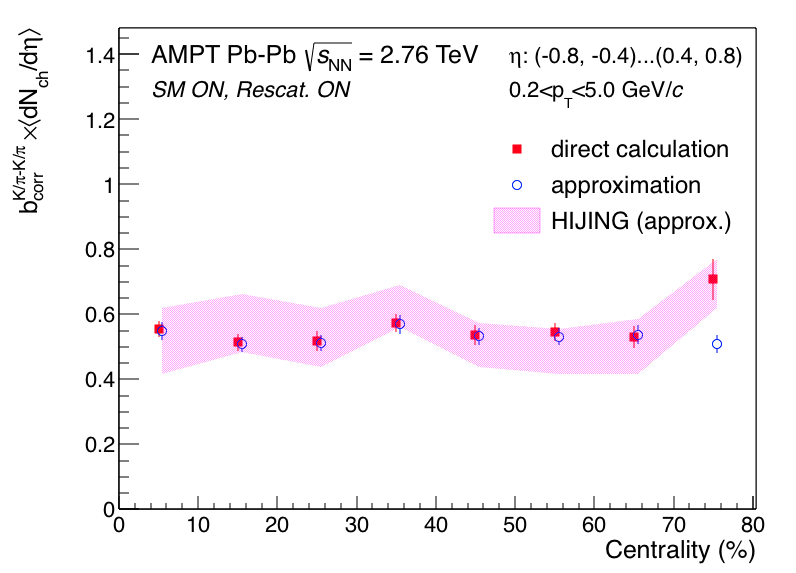}
\end{overpic}
\hspace{-0.1cm}
\begin{overpic}[width=0.48\textwidth]{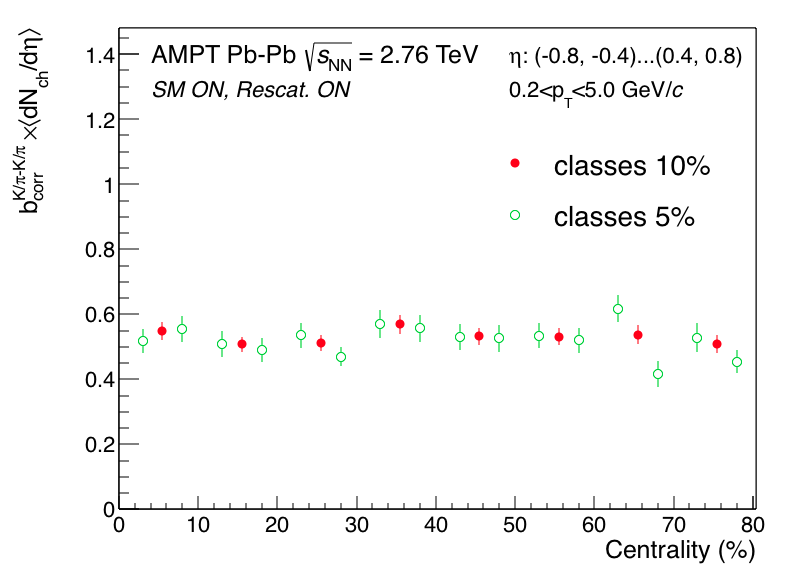}
\end{overpic}
\caption{
Scaled strength of forward-backward  correlations  between  $\nK/\npi$  ratios
as a function of centrality in Pb--Pb collisions at $\sNN=2.76$ TeV.
Forward-backward  $\eta$-intervals are ($-0.8, -0.4)$ and $(0.4, 0.8)$,
 $\pt$ range of particles is  0.2--5.0 GeV/$c$. 
{\it Left:}  direct calculations in AMPT (full squares)  in comparison with the approximation (open circles). Results from HIJING are shown as a filled band. 
{\it Right:} correlation strength in centrality classes of widths 10\% and 5\% (classes are determined 
using charge particle multiplicities at forward $\eta$-ranges
corresponding to the V0 detector of ALICE).
}
\label{fig:FB_rF_rB_K_pi} 
\end{figure}

Values of the correlation strength $\bcor^{\rm K/\pi}$  
 were calculated   
  in Pb--Pb collisions at $\sNN=2.76$ TeV 
in  AMPT  and HIJING  event generators.
 AMPT version with the string melting and rescattering options turned on was used.
On the left panel
in  Figure \ref{fig:FB_rF_rB_K_pi}, centrality dependence of the $\bcor$
calculated by the direct formula \eqref{corr_coeff}
 is shown in red squares.
Values are multiplied by charge particle density $\av{{\rm d}N_{\rm ch}/{\rm d}\eta}$ 
in order to cancel out the $1/\av{N_{\rm s}}$ scaling of this observable mentioned earlier.
Scaled  values  of $\nuFB$ given by  the approximation \eqref{beta_approx_using_R2_K_pi}
are shown as blue circles, 
 it can be seen that they are in agreement with the direct evaluation of $\bcor$. 
Centrality dependence of the correlation strength is flat in AMPT,
similarly to HIJING (shown on the same plot as a shaded band),
and the correlation strength is positive.
The right panel of the Fig. \ref{fig:FB_rF_rB_K_pi} demonstrates
robustness of this quantity to the volume fluctuations: 
values of $\bcor$ are the same for centrality classes of different width.

\begin{figure}[t]
\centering
\begin{overpic}[width=0.48\textwidth]{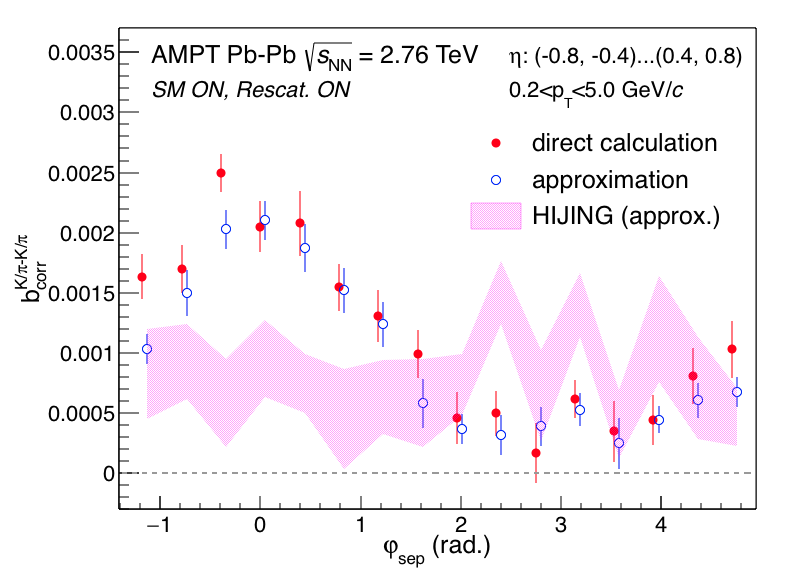}
\put(20,54){\footnotesize centrality 10-20\%}
\end{overpic}
\hspace{-0.1cm}
\begin{overpic}[width=0.48\textwidth]{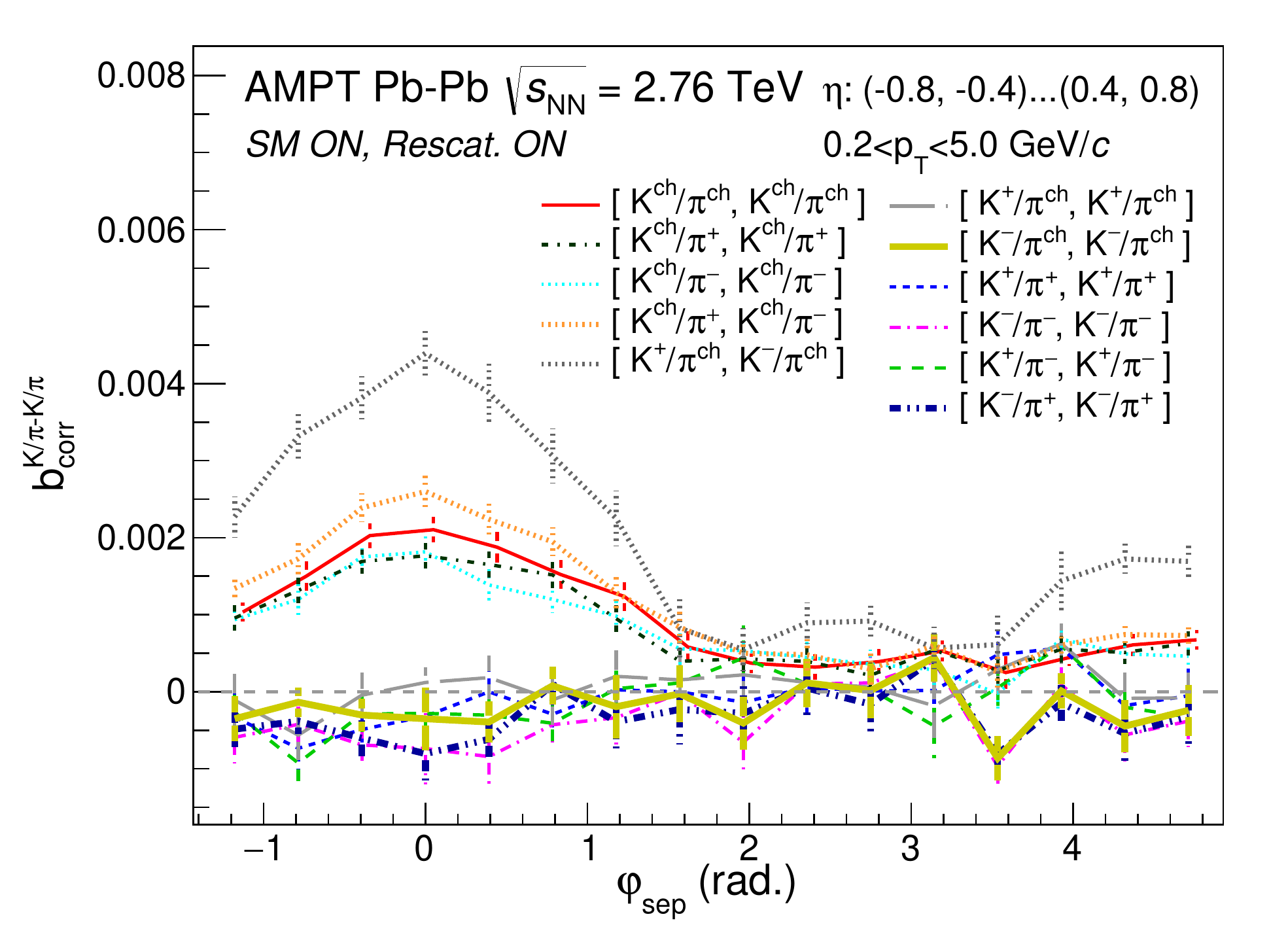}
\end{overpic}
\caption{
Forward-backward correlation strength $\bcor^{\rm K/\pi}$ as a function of the distance $\vf_{\rm sep}$ between the two azimuthal windows in 10-20\% centrality class of Pb--Pb collisions 
 in AMPT.
{\it Left:}  direct calculations in AMPT (full circles)  in comparison with the approximation (open circles). Results from HIJING are shown as a filled band. 
{\it Right:} correlation strength 
for different charge combinations of kaons and pions in the two windows.
}
\label{fig:FB_rF_rB_K_pi_phiWins} 
\end{figure}

The same analysis can be performed with the additional splitting of the $\eta$-windows 
into azimuthal sectors.
Left panel of Figure \ref{fig:FB_rF_rB_K_pi_phiWins} shows dependence of $\bcor$
on the distance $\vf_{\rm sep}$ between the two azimuthal windows in Pb--Pb collisions 
in  10-20\% centrality class in AMPT.
The approximate expression $\nuFB$ works  well even in this case,
when particle multiplicities in the acceptance windows are quite low 
(in each $\vf$-window here, $\av{\npi} \approx 13$, $\av{\nK}\approx 1.5$).
The ``near-side'' peak at $|\vf_{\rm sep}|<\pi/2$ is observed, while
$\bcor$ is close to 0 at $\vf_{\rm sep}\approx \pi$.
As for HIJING, results  seem to be consistent with a flat dependence on $\vf_{\rm sep}$,
however,
statistical uncertainties are quite large 
to make a definite statement.

Obviously, the correlation strength is affected by different contributions not directly related
to the ``collective effects'' we would like to access.
The main ``non-dynamical'' contributions are resonance decays and conservation laws. 
For example, in case of $\rm K/\pi$ analysis, the strangeness conservation
provides correlations which can be seen in the right panel 
of the Fig. \ref{fig:FB_rF_rB_K_pi_phiWins}:
the solid red line shows results for all charged kaons and pions measured in each window,
while
other lines correspond to more differential results,
when only certain charge combinations are used.
It can be seen, that when kaons of the opposite sign 
are counted in both forward and backward window (this corresponds
to the first column in the legend for this plot), 
the correlation strength is significant. 
The largest correlation, in fact, is for the case when one window counts
only K$^+$ and another -- only  K$^-$ (upper gray curve).
On the other hand, if only kaons of a  particular charge sign are counted in both windows
(right column in the legend),
the correlation strength is nearly zero.

\vspace*{-0.25cm}

\section{Correlations between particle ratio and mean transverse momentum}
\vspace*{-0.1cm}

Here we introduce  another potentially useful  observable, namely, 
a correlation between  particle yield ratio in one 
acceptance window and mean transverse momentum of particles in another:
\begin{equation}
\label{corr_coeff_rB_ptF}
\bcor^{\overline{p} -r}
= \frac{ \av{\pf\cdot \rB}}{  \av{\pf} \av{\rB} } -1 \, ,
\end{equation}
where $\overline{p} \equiv \sum_{i=1}^{n}\pt^{i}/n$  is the mean transverse momentum of particles
in a given event within acceptance cuts.
In case of modest event-by-event fluctuations 
of particle multiplicities with respect to their means,
  the approximate expression 
for this observable is 
\begin{equation}
\label{nu_FB_meanPt_r}
\bcor^{\overline{p} -r} \approx 
\frac{\av{\pf \cdot \nKB}}{\av{\pf}\av{\nKB}} 
-\frac{\av{\pf \cdot \npiB}}{\av{\pf}\av{\npiB}} 
\stackrel{\rm def.}{\equiv} \nuFB^{ \overline{p} -r } \, .
\end{equation}
Properties of this quantity,
namely, scaling with a number of particle sources
and robustness to efficiency and volume fluctuations,
 are similar to the $\nuFB$ introduced above.
With this observable,
one can study, for instance,
correlations between relative number of strange
particles 
in one rapidity interval and
a density of the fireball  (which reflects itself in $\overline{p}$)
 in another.

\begin{figure}[t]
\centering
\begin{overpic}[width=0.48\textwidth]{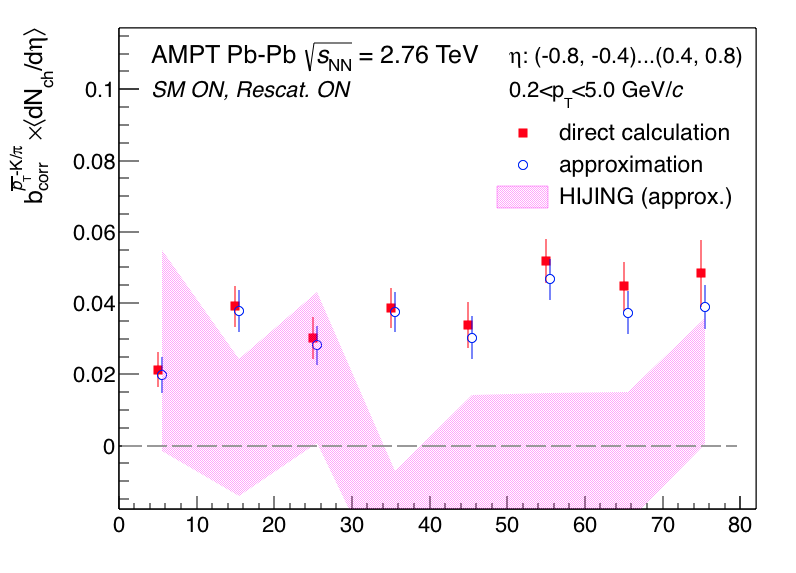} 
\end{overpic}
\hspace{-0.1cm}
\begin{overpic}[width=0.48\textwidth]{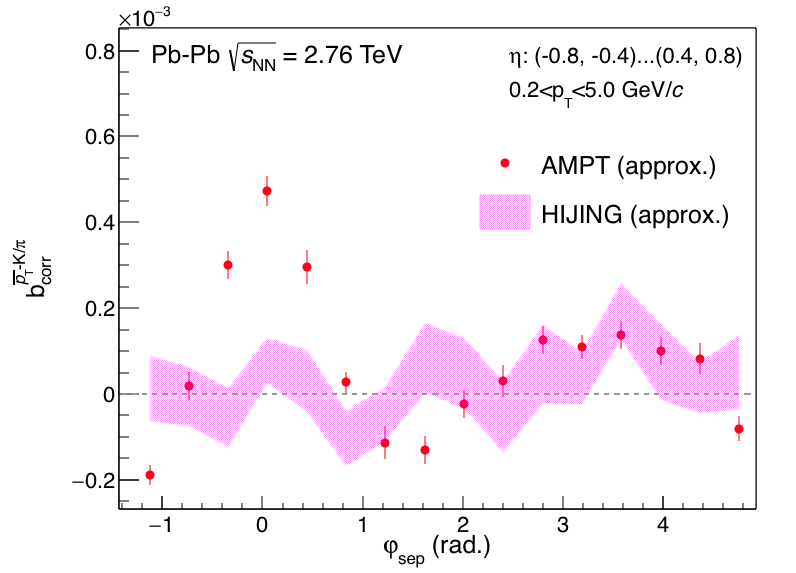}
\put(20,56){\footnotesize centrality 10-20\%}
\end{overpic}
\caption{
Forward-backward correlation strength 
$\bcor^{\overline{p} -{\rm K/\pi}}$
in Pb--Pb collisions  in AMPT
 as a function of collision centrality, scaled by charged particle density ({\it left} panel),
and as a function of the distance $\vf_{\rm sep}$ between the two azimuthal windows in 10-20\% centrality class ({\it right} panel).
Results from HIJING are shown as filled bands. 
Kinematic ranges are indicated on the plots.
}
\label{fig:FB_rB_ptF_K_pi} 
\end{figure}

Let us take for the backward window
the ratio between number of kaons to number of pions,
as before.
Left panel of Figure \ref{fig:FB_rB_ptF_K_pi}
shows centrality dependence of the $\bcor^{\overline{p} -{\rm K/\pi}}$
calculated in AMPT for the pair of $\eta$-windows, scaled with the multiplicity density.
The same plot demonstrates also the agreement between the approximate expression \eqref{nu_FB_meanPt_r} 
and the direct calculations.
On the right panel,
results of the analysis in azimuthal intervals
are shown as a function of $\vf_{\rm sep}$
(analogously to the Fig.~\ref{fig:FB_rF_rB_K_pi_phiWins}).
There is a peak at the near-side, 
and also some away-side 
structure is observed.
On both plots, values of $\bcor^{\overline{p} -{\rm K/\pi}}$ in HIJING 
are consistent with zero, which is in line with the absence of collective effects in this event generator.

Non-zero values of $\bcor^{\overline{p} -r}$ 
may provide new information about the collective effects 
in the medium. 
In order to further suppress contributions from resonances
and conservation laws, one could, in addition to 
the large $\eta$-gap between the windows, 
take for the ratio and mean $\pt$
only the same-charge particles, or differentiate the analysis in other ways, for example, calculate $\overline{p}$ only for  particles of a particular species.


\section{Summary and outlook}

In this article, 
we briefly reviewed 
a set of fluctuation-based observables 
which possess useful properties like robustness against limited particle registration efficiency
and  fluctuations of the collision geometry, 
and mentioned  some recent experimental measurements with these quantities 
in A-A collisions over the past few years.

In the second part of the paper, new  observables for fluctuation studies were introduced.
The first type of the proposed observables quantifies 
the correlation between 
ratios of identified particle yields measured
 in two separated acceptance  bins, 
the second type -- the correlation between the  ratio in one bin and 
average transverse momentum in the other.
With such observables it is possible, for instance, to study
a correlation between relative strangeness yield in one rapidity interval
and the density of the fireball, formed in A-A collisions, in another interval. 
Approximate expressions, 
which reveals a connection between the correlation strengths 
and a sum of normalized cumulants, were provided.
It was shown that these measures  
are also immune to volume fluctuations and detector inefficiencies.
The problem of particle mis-identification for these observables can be solved 
by utilizing the Identity Method. 
Evolution of the proposed quantities with centrality of A-A collisions
was calculated in AMPT and HIJING event generators.
Behaviour of these quantities in other models 
is to be investigated.


\section*{Acknowledgements}
This work is supported by the Russian Science Foundation, grant 17-72-20045.


\end{document}